\documentclass[aps,twocolumn,10pt]{revtex4}  
\usepackage{graphicx}
\usepackage[dvips]{color}
\input{epsf.tex}
\newcommand{\bibi}{\bibitem}                                                  
\newcommand{\etal}{\it {et al.}}                                              

\newcommand{\half}{\frac {1}{2}}                                             
                                             
\newcommand{\beq}{\begin{equation}}                                           
\newcommand{\eeq}{\end{equation}\noindent}                                  
\newcommand{\beqr}{\begin{eqnarray}}                                          
\newcommand{\eeqr}{\end{eqnarray}\noindent}

\newcommand{\vk}{{\bf k}}                                                     
                                                     
\newcommand{\vq}{{\bf q}}

\newcommand{\ad}{a^{\dag}}

\begin{document} 
\title{Spin-charge split pairing in underdoped cuprate superconductors: support from low-$T$ specific heat}

\author{Sanjoy K. Sarker}
\affiliation{Department of Physics and Astronomy, The University of Alabama, Tuscaloosa, Alabama 35487, USA}
\author{Timothy Lovorn}
\affiliation{Department of Physics and Astronomy, The University of Alabama, Tuscaloosa, Alabama 35487, USA}
\affiliation{Department of Physics, The University of Texas at Austin, Austin, Texas 78712, USA}
                       
\begin{abstract}

We calculate the specific heat of a weakly interacting dilute system of bosons 
on a lattice and show that it is consistent with the measured electronic
specific heat in the superconducting state of underdoped cuprates 
with boson concentration $\rho \sim x/2$, where $x$ is the hole (dopant) concentration.   
As usual, the $T^3$ term is due to Goldstone phonons. The zero-point energy, through its
dependence on the condensate density $\rho_0(T)$, accounts for the anomalous
$T$-linear term. These results support the split-pairing mechanism, in which spinons (pure spin)
are paired at $T^*$ and holons (pure charge) form real-space pairs at 
$T_p < T^*$, creating a gauge-coupled physical pair of charge $+2e$ and concentration
$x/2$ which Bose condenses below $T_c$, accounting for the observed phases.

   
\end{abstract} 
\maketitle                                                               
\vspace{0.5 in}                                                              



{\em {Introduction}}.---In
conventional superconductors electrons form spatially overlapping Cooper pairs below a temperature $T_c$. In contrast, the metallic phase of an underdoped cuprate superconductor with a small concentration $x$ of holes is not characterized by electron-like quasiparticles \cite{tim}. Instead, spin and charge appear to be carried by separate excitations, consistent with Anderson's proposal of spin-charge separation \cite{and}. For example, with decreasing $T$ the paramagnetic susceptibility decreases rapidly below a pseudogap temperature $T^* > T_c$, consistent with gapping of spin excitations, even as charge propagation becomes more metallic, as evidenced by the appearance of a Drude peak \cite{syr}. Nevertheless, the superconducting state is widely believed to be conventional, that is, a condensate of Cooper pairs with $d$-wave symmetry, 
primarily because of observations of electron-like peaks in the photoemission spectrum below $T_c$ {\cite{shen,shen2}. 

The conventional model, however, has serious inconsistencies. It cannot explain why pairs continue to exist up to $T_p$ far above $T_c$, but below $T^*$ \cite{ong1}. Since a Cooper pair has zero momentum a macroscopic condensate of such pairs is superconducting and hence should not exist above $T_c$. The problem is partly circumvented if electrons form nonoverlapping (real-space) pairs at 
a high temperature, with superconductivity occuring by Bose-Einstein condensation (BEC) \cite{chen}. However, cuprates are basically doped Mott insulators 
in which occupancy of a lattice site by two electrons is effectively forbidden
due to large on-site repulsion $U$. 
Then, given the high electron concentration ($1-x$),   
there is little room for creating nonoverlapping pairs. More importantly, it will be almost
impossible for two electrons to avoid double occupancy and propagate as a well-defined bound pair when
even a single electron has a hard time propagating with a well-defined momentum $\vk$. Indeed,
electron pairing cannot explain why electrons are not seen when pairs break in the metal. Or, why $T_c$ decreases with decreasing $x$, and superfluid density $\rho_s \sim x$ \cite{orn1}, (and not $1-x$), consistent with the carrier density ($x$) of the metal. 
It seems, therefore, that despite the appearance of the 
electron peaks below $T_c$, 
neither electrons nor electron pairs are the true low-energy excitations in the metallic or the superconducting phase.  Not suprisingly, a plausible microscopic theory connecting electron pairing with the actual excitations is yet to emerge.     

Recently a different, logically consistent, pairing mechanism has been found
in the spin-charge separated state of a {\em renormalized} $t$-$J$ model in 
which the spin and the charge of the electrons are paired separately \cite{sar9,sar10}. 
As an electron hops on the lattice with an amplitude $t$ avoiding double occupancy, 
its spin behaves like a localized moment, coupled with neighbors with an antiferromagnetic interaction $J$, as in the Mott insulator ($x = 0$). In the quantum antiferromagnet, 
spins are known to be paired into singlets which form a BCS-like, spin RVB state of paired spinons \cite{aro}. The charge ($+e$) is carried by a holon (concentration $x$), which is coupled to the spinon via a gauge field. The preformed pairs, Anderson argued \cite{and}, could lead to superconductivity for $x > 0$. But, (1) numerous attempts to construct an useful metallic state based on the bare $t$-$J$ model failed, and (2) the model does not have a mechanism to convert the chargeless RVB singlets into electron pairs. 
However, one-hole calculations \cite{kane} have shown that for $x > 0$, competition from AF correlation renormalizes the model with $t \gg J$ to one with $t_{eff} < J_{eff}$. 
This allows one to decouple the spinons perturbatively and obtain a {\em reduced} Hamiltonian \cite{sar9}, in which sublattice-mixing bare hopping $t$ is renormalized away and is replaced by
sublattice-preserving one-holon and holon-pair hopping terms accompanied by spinon singlet backflows, which cures both problems. 
The in-plane amplitudes for both one-hole and the pair hopping are the same: $t_s \sim 4t_{eff}^2/J_{eff} \propto J$, and $\ll t$.  
The new Hamiltonian is short-range RVB type \cite{krs}, with the pair hopping term supplying the missing {\em superconducting mechanism}.    

By requiring that the spin phases are the same (by symmetry) as those in the
insulator ($x = 0$), one obtains a qualitatively correct phase diagram for underdoped cuprates, as detailed in \cite{sar9,sar10}. Briefly, the phase above $T^*$ - the strange metal - has no quasiparticles of any kind. The spinon singlets condense below $T^*$ in the spin RVB state, accounting for the observed suppression of paramagnetic susceptibility. The RVB order
allows holons to propagate coherently and create a spinless Fermi liquid of concentration $x$ via the one holon hopping term, which is consistent with the observed transport properties \cite{syr,ando,pad} and quantum oscillation in a magnetic field \cite{tai}.  

The smallness of $x$ causes holons to form real-space pairs below $T_p < T^*$ via the strong coupling term \cite{sar10}. As it propagates, the phase of the bound (holon) pair is locked to that of the spin RVB ordered state, creating a physical (gauge invariant) pair with a split character.
It behaves like a mobile {\em hole} pair of charge $+2e$ with concentration $x/2$, 
which lives in a $d$-wave band; its spin part is an RVB singlet (concentration $1-x$), already ordered at $T^*$. This is seen in the calculated {\em electron pair} Green's function \cite{
sar10} 
$$ G^{el-pair}(\vk,\omega) = - \frac{A^2Z(-\vk)}{i\omega + \epsilon _b (-\vk)}.$$
where $A$ is the spin RVB order parameter, $Z$ is the pole strength, and $\epsilon_b (-\vk) \ge 0$ is the energy of the bound holons. Thus the mobile part of the electron pair actually has
negative energy and opposite momentum, i.e., it behaves like a hole pair. 
Superconducivity appears via BEC of these pairs below $T_c$, which 
accounts for the decline of $T_c$ with decreasing $x$ (for a noninteracting Bose gas $T_c \sim x^{2/3}$) and the fact that $\rho _s \sim x$.  

{\em {Specific heat}}.---In this Letter we calculate the low-$T$ specific heat $c(T)$, a bulk property which
directly reflects the dominant low-energy excitations in the superconductor, and thus
provides a serious test of the theory. In cuprates, the background, 
including the lattice phonon contribution, is removed, e.g., by subtracting 
specific heat measured in a modest magnetic field \cite{huen,marc}. Since $T^* \gg T_c$ the field does not break the singlets, but removes contributions from the (chargeless) spin (zero) sector. 
Then, at low $T$, we can model the system with weakly interacting bosons of concentration $\rho = x/2 \ll 1$ hopping on a lattice, and interacting via a weak short-ranged repulsion. 
Although cuprates are layered, low-$T$ physics for small $\rho$ can be described by scaling the wave vectors and using an isotropic 3D bosonic band of energy $\epsilon_b (k) = t_bk^2$, with  $t_b = (t_{ip}^2t_{op})^{1/3}$, where $t_{ip}$ and $t_{op} < t_{ip}$ are the in-plane and out-of-plane boson hopping parameters which can be calculated from the renormalized model \cite{sar10}. 

Above $T_c$ the measured $c(T)$ does not agree with the BCS theory. Instead, it declines slowly from a maximum as in a Bose gas, 
with an eventual crossover to metallic ($T$-linear) behavior above $T_p$ \cite{huen}, 
as expected in our model. Moreover, the maximum (and the overall scale) increases with $x$, not $1-x$ \cite{loram}. For $T \ll T_c$, the measured $c(T)$ in many cuprates behaves like \cite{marc}
\beq c(T) = AT + BT^3. \eeq
If density of states of the relevant excitations behaves like $D(\epsilon) \propto \epsilon^p$, 
for small excitation energy $\epsilon(q)$, then $c(T) \propto T^{p+1}$. For the $d$-wave BCS state, the excitations are the nodal electrons with $p = 1$, and the absence of a $T^2$ term in c(T)
implies that the electron {\em peaks} in the photoemission spectrum do not behave like {\em long-lived quasiparticles}, effectively ruling out this state. 

For phonons $\epsilon (q) \propto q$, giving $p =2$, and $c(T) \propto T^3$. However, a definite curvature in $\gamma = c/T$ remains even after lattice phonons are removed,  
indicating that $T^3$ term in (1) is of electronic origin. For a dilute Bose gas, Bogoliubov has shown that \cite{bog}, coupling of bosons with the condensate leads to a
depleted condensate density $\rho_0(T)$, and Goldstone phonons, which yields a $T^3$ specific heat \cite{yang}. While this agreement with Eq. (1) is encouraging, the $T$-linear term poses a fundamental challenge, since it dominates at small $T$, and it has become increasingly clear that it cannot be 
attributed to disorder \cite{marc}.

{\em {Weakly interacting bosons, low $T$}}.---This
leads us to re-examine the Bogoliubov theory at low $T$. The point is, in addition to the direct ($\sim T^4$) contribution to the energy from Bogoliubov phonons, there is also a zero-point energy which depends on $T$ through its dependence on $\rho_0(T)$, and thus contributes a $T$-linear term to the specific heat. To see this, consider the Hamiltonian       
\beq
H = \sum _{\vk} \epsilon (\vk) \ad _{\vk}a_{\vk} +  \frac{1}{2N} \sum _{\vq,\vk_1,\vk _2} V(\vq)
	\ad _{\vk _1 +\vq}\ad _{\vk _2 - \vq}a_{\vk _2}a_{\vk _1}, \eeq
where $\epsilon _b(\vk) = t_b k^2$, $V(q)$ is a weak repulsive
interaction, and $a_{\vk}$ destroys a boson of momentum $\vk$. 
For uniform $\rho$, the Hartree ($V(0)$) term is a constant and is subtracted out. We also approximate $V(q)$ by a constant which does not change the results qualitatively. Symmetry is broken below $T_c$, so that $\langle a_0 \rangle = (N\rho_0)^{\half} \ne 0$. Replacing $a_0$ by its average and retaining quadratic terms we obtain the Boguliobov Hamiltonian 
\beq
H_{B}(\rho_0) = \sum _{\vk \ne 0} [(\epsilon (\vk) + \Sigma - \mu) \ad _{\vk}a_{\vk} + \frac{1}{2}\Lambda	 
(\ad _{\vk}\ad _{-\vk} + h.c.)], \eeq
where $\Sigma = \rho_0V$ and $\Lambda (k) = \rho _0V$ are self energies, $\mu$ is the chemical potential. The partition function is calculated by using $H_B$, which depends on $\rho_0$ via $\Lambda$. Diagonalization leads to
$$ H_{B}(\rho_0) = U_0(\rho_0) + \sum _{\vk \ne 0} E(k)\alpha^{\dag}_k\alpha_k. $$
where $U_0$ is the zero point or \lq\lq vacuum" energy, which arises due to the anomalous coupling, and $\alpha ^{\dag}_k$ creates an excitation (out of this vacuum) of energy $E(k)$, which depends on $\rho_0$. In the Bogoliubov theory \cite{bog,FW} the spectrum is taken to be gapless
which gives $\mu = \Sigma - \Lambda = 0$ and
$E(k) = [2\Lambda\epsilon (k) + \epsilon^2(k)]^{1/2}$.   

For a given $\rho _0$ and $T$, the average energy ($k_B =1$) is then 
\beq U = U_0(\rho _0) + \sum _{\vk \ne 0}\frac{E(k)}{(e^{E/T} - 1)} = U_0 (\rho_0) + U_1(\rho_0,T), \eeq
where $U_1$ is excitation contribution which vanishes as $T^4$, yielding a $T^3$ term in $c$. 
$U_0(\rho_0)$ is the ground-state energy, which is usually taken to be a constant. However, since $\rho_0$ represents the condensate it is a thermodynamic variable, and thus depends on $T$, as emphasized by Yukalov \cite{yuka}.
It is obtained from the thermodynamic relation  
\beq \rho _0  = \rho - \rho_1, \eeq 
where $\rho_1 = \sum _{k \ne 0} n_k/N$ is the density of noncondensate bosons, with $n_k$ = $\langle a^{\dag}_ka_k \rangle$, calculated as a function of $T$ and $\rho_0$ by using $H_B(\rho_0)$. Solving Eq. (5) self-consistently one obtains $\rho_0(T)$ and hence thermodynamic properties as a function of $T$ and $\rho$. The true ground state energy is given by $U(\rho_0(0))$. However, the behavior for $T > 0$
is not determined by the excitations about this $(T =0)$ vacuum, but by the true
many-body states in the ensemble characterized by $\rho_0$ at this $T$ (i.e, approximately by
those of $H_B(\rho_0(T)$); $U_0(\rho_0(T))$ is the ground state energy for this ensemble and hence
will contribute to $c(T)$. This is a general result which would apply to other systems 
since it is a consequence of thermodynamics. In particular, using Bogoliubov low-$T$ result $\rho_0(T) \approx \rho_0(0) - aT^2$ (where $a$ is positive constant) in  Eq. (4) one immediately obtains a $T$-linear term for $c(T)$. It arises from a nonperturbative change in phonon speed (not its $q$ dependence) in the presence of the condensate. The self-consistent procedure has been used in earlier $T > 0$ studies of a dilute Bose gas~\cite{yuka,griff},  but the effect on $U_0$  apparently has not been recognized. 

{\em {Hartee-Fock corrections}}.---Although
the Bogoliubov approximation agrees qualitatively with $c(T)$ at low $T$, it has
well-known problems \cite{anders}, particularly at higher-$T$, predicting a spurious first-order transition. This is corrected if the remaining interactions are treated by a self-consistent Hartree-Fock approximation. Our treatment is similar to earlier ones \cite{yuka,griff}, except that we calculate $c(T)$ and  in our lattice problem the bandwidth $W$ acts as a natural cutoff. 
We also solve the HF problem analytically at low $T$ which
allows as to obtain the crossover scale as a function of $\rho$ below which the linear term becomes
important.

The HF decomposition leads to the same quadratic Hamiltonian (3) except now, 
$\Sigma = V(\rho _0 + \rho _1) = V\rho$, and 
\beq \Lambda = V(\rho _0 + \sigma) = V(\rho - \rho _1 + \sigma), \eeq
where $\sigma = \sum _k \sigma _k/N$ is the anomalous density. It arises as a response 
to $\rho_0$, i.e., vanishes if $\rho_0$ does. Once $\mu$ is fixed by requiring gaplessness, the Hamiltonian depends on $\rho_0$ through  $\Lambda$. These are determined as functions of $\rho$ and $T$ by solving Eqs. (5) and (6) self consistently. 

As before, each of $\rho _1$, $\sigma$, and the energy per site $u = \langle H \rangle/N$ splits into vacuum and 
excitation terms; the latter depends on $T$. 
For example, $\rho _1  = \rho_{10}(\Lambda) + \rho _{11}(\Lambda,T)$.
For small $\rho$, the relevant energy scales $(\Lambda,T) \ll W$.  
The integrals are evaluated
to leading orders in $(\Lambda/W)$ etc. We set $\int_0^{W} d\epsilon D(\epsilon) = 1$, where $D(\epsilon) = \epsilon ^{1/2}/\tau^{3/2}$ is the density of states, with $\tau = (4\pi^2)^{2/3}t_b$,  which gives $W = (3/2)^{2/3}\tau$. 
At low-$T$, HF theory provides a screened interaction 
$ g = V/[1 + VW^{1/2}/\tau ^{3/2}]$. Then the solution at $T = 0$ is 
\beq \Lambda (T=0) = \Lambda _0 \approx \rho g(1 + \eta/3), \eeq
where $\eta = (2\Lambda _0/\tau)^{3/2}/\rho \approx (2g/\tau)^{3/2}\rho^{\half}$,
emerges as the small parameter with the same $\rho ^{\half}$ dependence as in the Bogoliubov
theory. For $V \sim t$ and $\rho <0.1$, $\eta < 0.023$ is very small. The fractional depletion of the condensate is given by $\rho _1/\rho  \approx \eta/6$.    

The characteristic energy scale at low-$T$ is $\Lambda_0$.  
The energy per site is $u = u_1 + u_2$, where $u_1 = \sum _k \epsilon (k)n_k/N$, and the average interaction energy $u_2$ is given by
\beq u_2 = \frac{V\rho ^2}{2} + \frac{\Lambda ^2}{2V} - V\rho_0^2. \eeq 
At low-$T$, we have solved the consistency equations by expanding 
in powers of $(T/2\Lambda_0)^2$, and obtaining $u$ up to $(T/2\Lambda_0)^4$. 
This gives
\beq c = \rho\eta\frac{\pi^2}{3g}\frac{T}{2\Lambda_0}[(V - g) + \frac{2\pi^2}{5}(5g- V)(\frac{T}{2\Lambda_0})^2 + O(T^4)],\eeq 
which is valid for $\Lambda _0 \approx \rho g \gg T$, and has the observed form (Eq. (1)). Thus the coefficient of the linear term $A \propto \rho ^{1/2}$, and of the cubic term $B \propto \rho ^{-{3/2}}$. 
 
The crossover scale $T_{cr}$ above which the $T^3$ becomes larger than the linear term is
$T_{cr} = \Lambda _0 [(10/\pi^2)(V-g)/(5g-V)]^{1/2}$. It is useful to compare $\Lambda _0$ and $T_{cr}$ with $T_c$. Above $T_c$, the Hartree and exchange terms are constant and $\rho _0 = 0 =\sigma$.
Hence, $T_c$ is given by the noninteracting expression $T_c/\tau = [2\rho/(\sqrt{\pi}\zeta(3/2))]^{2/3}$, where $\zeta(z)$ is the Riemann zeta function. Then, for small $\rho$, $\Lambda _0/T_c \approx  0.151 (g/t_b)\rho ^{1/3}$, which is rather small. And for $V \sim t_b$ or less, 
$T_{cr}/T_c \approx 0.024 (V/t_b)^{3/2}\rho ^{1/3}$, where $0.024$ is a geometrical factor.
Since $T_{cr}/T_c$ is so small finding the linear term experimentally in very low $T_c$ Bose systems such as atomic gases would require considerable care. 
The result for $T_{cr}$ will be different for a strongly interacting, high density system like
liquid He$_4$, in which phonons exist even above $T_c$, but the effect should exist.

{\em {Near $T_c^-$}}.---As
$T \rightarrow T_c$ from below, the densities show a square root singularity as $\Lambda \rightarrow 0$, e.g., 
\beq \rho _{11}(\Lambda,T) =  \rho (T/T_c)^{3/2} +
(\frac{1}{\tau})^{\frac{3}{2}}[- a_1\Lambda ^{\half}T + a_2 \Lambda T^{\half}],\eeq 
where $a_1 = \pi/\sqrt 2$ and $a_2 = - {\sqrt \pi}\zeta (1/2)/2$ with $\zeta (1/2) = - 1.4604$.
The zero-point term $\rho _{10} = (\Lambda/\tau)^{3/2}/6$ is of higher order. 
Without the HF correction $\Lambda = V\rho_0$, 
then the dominant $\Lambda ^{1/2}$ term ensures that Eq. (5) has no {\em real} solution for $\rho _0$ close to $T_c^-$, resulting in a first-order transition. 
This is corrected in the HF theory since the 
anomalous self-energy $\sigma$ has the same square-root term which cancels the one
of $\rho _1$ in Eq. (6). This follows from the identity $(\partial \rho_{11}/\partial \Lambda)_T = \sigma _1/2\Lambda$ satisfied by the $T$-dependent parts $\rho _{11}$ and $\sigma _1$. )Solving Eq. (6) we obtain 
$$ \Lambda \approx g_2\rho (1 - (T/T_c)^{3/2}) \sim (g_2/T_c)(T_c - T),$$
where $g_2 \approx g/[1 - ga_2 T_c^{\half}/\tau ^{\frac{3}{2}}]$.
Then
$$\rho _0 = \rho - \rho_1 \approx \rho(1 - (T/T_c)^{3/2}) + a_1\frac{\Lambda ^{\half}T}{\tau ^{3/2}} \propto (T_c-T)^{\half}. $$ 
Thus the transition is second order and with the usual mean-field exponent ($\half$).
Using another identity $(\partial u_{11}/\partial \Lambda)_T = - \frac{3}{2} (\rho _{11}+\sigma _{11})$, we obtain the energy 
$$u =  \half V\rho ^2  - V a_1^2 \frac{T^2\Lambda}{\tau^3}   + \frac{3}{2}\rho(T/T_c)^{\frac{3}{2}}[\frac{\zeta(5/2)}{\zeta(3/2)}T-\Lambda],$$ to order $\Lambda$, from which we obtain the specific
heat
$c = du/dT$,
$$c \approx c_B - (\frac{2Va_1^2T}{\tau^3} + \frac{9\rho}{4T_c}(\frac{T}{T_c})^{\half})\Lambda -(\frac{Va_1^2T^2}{\tau^3}+ \frac{3\rho}{2}(\frac{T}{T_c})^{\frac{3}{2}})\frac{d\Lambda}{dT}, $$
where
$c_B = \frac{15}{4}\frac{\zeta(5/2)}{\zeta(3/2)}\rho(T/T_c)^{\frac{3}{2}}$ 
is the specific heat per site of nointeracting bosons.  
As is usual in a mean-field theory, $c(T)$ has a jump discontinuity at $T_c$
$$\frac{c - c_B}{c_B} = \frac{2}{5}\frac{\zeta(3/2)}{\zeta(5/2)}\frac{g_2}{\tau}[\frac{3}{2}\nu^{2/3}\rho^{1/3} + \frac{Va_1^2}{\tau}(\rho/\nu)^{2/3}],$$
where $\nu = \Gamma (3/2)\zeta(3/2)$. Thus the fractional jump is nonuniversal and vanishes as   
$\rho ^{1/3}$. At $T_c$, $c(T_c) \propto \rho ^{4/3}$. An increase with $\rho$ has been observed, although the power is unknown. The HF results are a major improvement, but obtaining the true critical behavior would require
an RG treatment, and additionally gauge fluctuations may have to be included.    

{\em {Conclusions}}.---Split
pairing with real-space pairing of holons, as described in \cite{sar9,sar10} 
is thus consistent with the observed specific heat, 
superfluid density, decline of $T_c$ with decreasing $x$, 
and the existence of pairs above $T_c$ as inferred from the observed Nernst effect, diamagnetism, and specific heat.
The crossover to a spinless Fermi liquid 
of carrier charge $+e$ and concentration $x$ is consistent with the observed properties in the pseudogap metal, including quantum oscillations. 
Pairing of spinons (concentration $1-x$) below $T^*$ describes the Mott physics, and gauge coupling with holons accounts for the existence of two metallic phases 
and connects the spin sector to the insulator through RVB ordering.    

Why does the nodal BCS theory based on the electron peaks observed below $T_c$ fail 
to describe the low-$T$ properties such as specific heat? The plausible answer is that 
these are not long lived, but break up into true long-lived  spin and charge excitations.
The calculation of the electron Green's function
is difficult, and is not attempted here. Instead, we discuss the behavior qualitatively. 
In a spin-charge separated state an electron of momentum $\vk$ decays
into a spinon and an (anti)holon, subject to energy and momentum conservation. 
The spectral weight is distributed over all such pair states
in the spinon-holon continuum. Hence, while there is a weak peak, there is no pole ($Z = 0$) and
the width is of order of bare bandwidths $\sim t$ \cite{sar12}, as seen. About $T^*$ and below, the low-energy region is described by the renormalized Hamiltonian, characterized by a smaller bandwidth $\sim t_s$, the renormalized hopping parameter which is $\sim J_{eff} 
\ll t$. Once spinons are paired and then holons, it costs energy to free them, 
making some decay channels less probable, allowing electrons to propagate longer.
This is accentuated below $T_c$, as a macroscopic fraction of spinons and holons
remain in the corresponding condensates, so that electrons propagate even longer. 
It is not surprising that electron peaks appear, but these additional structures are also broad,
with width $\sim t_s$, the low-energy bandwidth. 

This is borne out by the observed photoemission spectra \cite{shen,shen2} which show that the width of a single electron peak (at $\vk$) is as large as $\sim 20$ meV at half maximum, and at the base it is $\sim 0.1$ eV, i.e., of order the smaller bandwidth. 
In other words, the spectral weight of each electron is spread over the entire bandwidth of the {\em renormalized} particles. The electron excitations are thus short lived resonances even in the superconducting state, not true quasiparticles.    

\begin{acknowledgments}
This research was supported by the National Science Foundation (DMR-1508680).
Work in Austin was supported by the Department of Energy, Office of Basic Energy Sciences under contract DE-FG02-ER45118 and by the Welch Foundation under grant TBF1473.
\end{acknowledgments}

\end{document}